\documentstyle[aps,epsf]{revtex}
\def\vec#1{\mathchoice
{\mbox{\boldmath $\displaystyle#1$}}
{\mbox{\boldmath $\textstyle#1$}}
{\mbox{\boldmath $\scriptstyle#1$}}
{\mbox{\boldmath $\scriptstyle#1$}}}

\newcommand{\ket}[1]{| #1 \rangle}

\newcommand{\element}[3]{\langle #1 | #2 | #3 \rangle}

\newcommand{\G}{\mbox{$\Gamma_{pair}\,$}}
\newcommand{\D}[1]{\mbox{$\Delta_{#1}$}}
\newcommand{\r}[1]{\mbox{$\rho_{#1}$}}
\newcommand{\k}[1]{\mbox{$\tilde{\rho}_{#1}$}}

\begin{document}

\title{
Hartree-Fock-Bogoliubov calculation of charge
radii of Sn, Ba, Yb, and Pb isotopes
}

\author{
Shouichi Sakakihara
\footnote{{\rm E-mail shouichi@npl4.kyy.nitech.ac.jp}},
Yasutoshi Tanaka
\footnote{{\rm E-mail tanakay@ks.kyy.nitech.ac.jp}}
\\[1ex]
\it
Department of  Environmental Technology and Urban Planning, \\
\it
Nagoya Institute of Technology,
Gokiso, Nagoya 466-8555, Japan
}

\maketitle

\begin{abstract}

Charge radii of Sn, Ba, Yb, and Pb isotopes are calculated
within Hartree-Fock-Bogoliubov theory with a Skyrme force
and a density-dependent delta-force pairing.
We investigate mean field effects of the pairing upon
odd-even staggering of isotope shifts.
HFB equations are solved in the canonical basis.
Odd nuclei are treated in the blocking approximation.
\end{abstract}

\section{INTRODUCTION}
{\label{secINT}}

Odd-even staggering of isotope shifts is a common
phenomena observed in many isotopic chains {\cite{AH87}}. 
The results are interpreted as due to changes 
in the charge radii of these isotopes, i.e., charge
radii of odd-neutron nuclei are smaller than
the average radii of their even neighbors {\cite{AH87}}.

Probable mechanisms so far presented 
for odd-even staggering
of isotope shifts are; 
core polarization by valence neutrons {\cite{IT84,SA87}}, 
and mean field effects of 
a three-body (or a density-dependent) pairing force 
{\cite{{RZ88,FT94,KT95,FZ96,TB93}}}. 
Core polarization effects were first investigated by Talmi
{\cite{IT84}} as a possible mechanism for producing
odd-even staggering of isotope shifts.
He could show a remarkably good fitting to changes in the charge
radii of Ca isotopes, though it is not certain whether
the shell model calculation with a realistic force
can reproduce absolute magnitude of isotope shifts.
An alternative mechanism was suggested by Zawischa et al.
{\cite{RZ88,FT94,KT95,FZ96}}.
They have demonstrated the importance of a density
dependent pairing as well as mean field effects of the
pairing on the odd-even staggering of isotope shifts.
They could obtain good fitting to changes in the charge
radii of Ca, Ba, Sn, and Pb isotopes.

In this paper we investigate in detail the latter possibility
employing the HFB theory with axially-symmetric deformation.
We assume the Skyrme SLy4 parameterization{\cite{CB98}}
and the delta-force pairing with linear density dependence
{\cite{TB93}}.
We solve HFB equations in the canonical basis{\cite{RB97,T98}}. 
Odd nuclei are treated in the blocking approximation.

This paper is organized as follows. In Section 2, 
we review some theoretical formalism useful for our purposes. 
In Section 3, we give numerical results and discussions 
on Sn, Ba, Yb, and Pb isotopes.
Summary and conclusions are given in Section 4.

\section{FORMALISM}
{\label{secFOR}}

We follow the HFB scheme prescribed by Reinhard et al
{\cite{RB97}}.
We start with the Hamiltonian,

\begin{equation}{\label{Hamiltonian}}
H = T + V^{Skyrme} + V^{pair} , 
\end{equation}

\begin{equation}{\label{DDDI}}
V^{pair}(1,2) = V_0 \frac{1 - \vec{\sigma}_1 \cdot \vec{\sigma}_2}{4}
                 \delta(\vec{r}_1 - \vec{r}_2) 
                 \left(
                 1 - \frac{\rho(\frac{\vec{r}_1 +
                            \vec{r}_2}{2})}{\rho_c}
                 \right) .
\end{equation}
Here $V^{Skyrme}$ is the Skyrme force and $V^{pair}$
is the density-dependent delta-force pairing {\cite{TB93}}.
The pairing force involves two parameters, i.e.,
$V_0$ sets the strength of the force
and $\rho_c$ determines its spatial dependence.

The BCS ansatz for the wave function of a pairing many-body system

\begin{equation}
\ket{\Phi_{BCS}} = \prod_{i > 0} 
  (u_i + v_i {a_i}^{\dagger} {a_{\bar{i}}}^{\dagger}) \ket{-}
\end{equation}
requires that the single particle states are orthonormalized
and that occupations add up to the total particle number
$\Sigma v_\alpha^2=N$. 
This ansatz carries two key densities {\cite{DF84}}. 
One is the one-body density,
\begin{equation}
\rho_q(\vec{r})   = \sum_{\sigma} 
                    \element{\Phi_{BCS}}
                            {{a^{\dagger}_{\vec{r} \sigma q}} 
                              a_{\vec{r} \sigma q}}
                            {\Phi_{BCS}}
                  = \sum_i v_i^2 \sum_{\sigma}
                    \phi_i(\vec{r},\sigma,q)
                    \phi_i^*(\vec{r},\sigma,q) ,
{\label{density}}
\end{equation}
and the other is the pairing density,
\begin{equation}
\k{q}(\vec{r}) = \sum_{\sigma} (-2 \sigma)
                    \element{\Phi_{BCS}}
                            {{a_{\vec{r} - \sigma q}} 
                              a_{\vec{r}   \sigma q}}
                            {\Phi_{BCS}}
               = \sum_i u_i v_i \sum_{\sigma}
                  \phi_i(\vec{r},\sigma,q)
                  \phi_i^*(\vec{r},\sigma,q).
\label{pairing density}
\end{equation}
Here $\vec{r}$, $\sigma$, and {\it q} denote the
space, spin, and charge state of a nucleon, respectively.

We assume the Skyrme force in the particle-hole channel, 
while the pairing force in the particle-particle channel. 
Thus the energy separates into a mean-field part and a pairing part as
\begin{equation}{\label{Total Energy}} 
E[\phi,v] = \element{\Phi_{BCS}}{H}{\Phi_{BCS}} 
= E_{HF}[\phi,v] + E_{pair}[\phi,v].
\end{equation}
$E_{HF}$ is the Skyrme Hartree-Fock energy and 
$E_{pair}$ is the pairing energy,

\begin{equation}{\label{pairing energy}}
E_{pair}[\phi,v] 
= \frac{V_0}{4} \sum_q \int \k{q}^2(\vec{r}) 
  \left( 1 - \frac{\rho(\vec{r})}{\rho_c} \right) d \vec{r}  .
\end{equation}

Variation of the energy {\it E} 
with respect to a single-particle wave function is written as

\begin{equation}{\label{variation1}}
 \frac{\delta E}{\delta \phi_i^*} = \hat{{\cal H}}_i \phi_i  ,
\end{equation}
\begin{equation}{\label{S.D.hamiltonian}}
\hat{{\cal H}}_i = v_i^2(\hat{h}_{Skyme} + \Gamma_{pair}(\vec{r})) 
                       + u_i v_i \D{q} (\vec{r})  .
\end{equation}
Variation of $E_{HF}$ yields 
the Skyrme Hartree-Fock hamiltonian $\hat{h}_{Skyrme}$ ,
while variation of $E_{pair}$ 
yields the pair mean-field \G 
as well as the gap potential \D{q}.
The pair mean-field \G comes from the variation of the density
$\rho_q(\vec{r})$ in Eq.~(\ref{pairing energy}).
It has the form

\begin{equation}{\label{Gammapair}}
\G(\vec{r}) 
    = - \frac{V_0}{4 \rho_c} \left(
        \k{n}^2(\vec{r}) + \k{p}^2(\vec{r})
        \right)  .
\end{equation}
The gap potential \D{q} comes from the variation of the pairing
density $\tilde{\rho}_q(\vec{r})$ in Eq.~(\ref{pairing energy}).
It has the form

\begin{equation}{\label{Gapfield}}
\D{q}(\vec{r}) 
    = \frac{V_0}{2} \k{q}(\vec{r})
      \left( 1 - \frac{\rho(\vec{r})}{\rho_c} \right)  .
\end{equation}

It is noted that the pair mean-field \G 
does not depend on the charge state.
It is always repulsive and arises only from the density dependent pairing.
On the other hand, the gap potential \D{q} 
depends on the charge state {\it q} and arises from any pairing force
irrespective of its density dependence.
For a density parameter \r{c} lower than the central density,
the gap potential will be repulsive inside the nucleus and
it is attractive at the nuclear surface.

         A single particle hamiltonian ${\cal {H}}_i$ of 
Eq.~(\ref{S.D.hamiltonian}) depends on the state {\it i}.
It is not hermitian in contrast to HF + BCS calculations.
Therefore variation actually needs to take care of the 
orthonormality of the wave functions as

\begin{equation}{\label{variation2}}
\frac{\delta}{\delta \phi_i^*} 
\left( E - 
\sum_{ij} \Lambda_{ij} \left( \int \phi^{*}_j (\xi) \phi_i(\xi) d\xi
 -  \delta_{ij} \right) \right)
= 0  ,
\end{equation}
where $\xi$ represents all variables.
The equation leads to a generalized mean-field equation,

\begin{equation}{\label{H.F.eq}}
\hat{{\cal H}}_i \phi_i = \sum_j \Lambda_{ji} \phi_j  .
\end{equation}
Lagrange multipliers $\Lambda_{ij}$
constitute a symmetric matrix 
$\Lambda_{ij} =  \Lambda_{ji}$ .

	Variation of the energy with respect to the occupation
probability under the particle number constraint yields

\begin{equation}{\label{BCSeq}}
\frac{d}{d v_i} \left( E - \lambda \sum_j v_j^2 \right) 
= 2 v_i h_{ii} + 
(\D{ii})_q \left(\frac{v_i^2}{u_i} - u_i \right)
= 0  ,
\end{equation}
\begin{equation}
h_{ii} = \element{\phi_i}
         {\hat{h}_{Skyrme} + \Gamma_{pair}(\vec{r}) - \lambda}
         {\phi_i}  ,
\end{equation}
and the pairing gap is calculated as

\begin{equation}{\label{Gap}}
(\D{ii})_q  = - \int \sum_{\sigma} |\phi_i(\vec{r},\sigma,q)|^2 
                  \D{q}(\vec{r}) d\vec{r}  .
\end{equation}
Eq.~(\ref{BCSeq}) can be solved in the standard manner and yields

\begin{equation}
           \left\{ \begin{array}{c}
                    v_i \\ u_i
                    \end{array}   \right\}
=   \sqrt{ \frac{1}{2} \mp 
           \frac{1}{2}
           \frac{h_{ii}-\lambda}
                {\sqrt{(h_{ii}-\lambda)^2+\D{ii}^2}} }
 . {\label{uv}}
\end{equation}
Both Eqs.~(\ref{H.F.eq}) and (\ref{BCSeq}) constitute
the HFB equation.
Eqs.~(\ref{H.F.eq}) and (\ref{BCSeq}) are
solved iteratively employing the gradient method,
\begin{equation}{\label{New vector}}
\ket{\phi_i^{n+1}} = {\cal O} \{
\ket{\phi_i^n} - 
{\cal D} (\hat{{\cal H}}_i^n \ket{\phi_i^n} - \sum_j \Lambda_{ji}
\ket{\phi_j^n})  \}.
\end{equation}
The parameter ${\cal D}$ determines the size of the step
and $\cal O$ means Gram-Schmidt orthonormalization.
$\ket{\phi_i^n}$ and $\hat{{\cal H}}_i^n$ 
represent a single particle state and the mean field 
in the n-th iteration, respectively. 
Lagrange multipliers $\Lambda_{ij}$ are 
determined by the orthonormality of the states.
They are
\begin{eqnarray}{\label{Lambda}}
\Lambda_{ij} &=& 
\frac{1}{2} \element{\phi_i^n}
                    {\hat{{\cal H}}_i^n + \hat{{\cal H}}_j^n}
                    {\phi_j^n}  .
\end{eqnarray}

\section{Results and discussions}
{\label{secRES}}

We have investigated changes in the charge radii of 
Sn, Pb, Ba, and Yb isotopes because
long isotopic chains are observed in these nuclei and
because they lie in the spherical, transitional and well deformed
regions of the nuclear chart, respectively.
The parametrization SLy4{\cite{CB98}} is 
employed for the Skyrme force.
The parameters were specifically devised 
to reproduce correct incompressibility modulus 
for the symmetric nuclear matter and to improve 
the isospin property of the force away from 
the $\beta$-stability line.

These features of the SLy4 force are especially favorable
to the calculation of charge radii along a long isotopic chain.
However, the SLy4 paramaters were determined 
without the pairing correlation
and therefore they would have to be
modified when they are used in HFB calculations.
With the pairing parameters discussed in the next paragraph,
it turns out that pair mean-fields \G
contribute to HF + BCS results of Sn, Pb, Ba, and Yb 
isotopes by an order of ten keV
and the gap potentials \D{} by an order of hundred keV.
Gap potentials have non-negligible effects on the
odd-even mass difference of isotopes.
However, the effects may be included in the pairing 
by adjusting its parameters and therefore
we assume the original SLy4 parametrization in the following analysis.

The depth and the density parameters of the 
pairing force are taken as $V_0$ = -- 1250 MeV$\cdot$fm$^3$
and $\rho_c$ = 0.140 fm$^{-3}$, respectively. 
They were obtained by
fitting to one-neutron separation enegies as well as
odd-even sttagering of charge radii of the above isotopes.
We assumed the same set of pairing parameters for all isotopes.
With a fixed value of \r{c}, we determined the depth $V_0$ 
by fitting to one-neutron separation energies.
We repeated this process for several different values of \r{c}
and found the best value of \r{c} by comparing
odd-even staggerings of charge radii.
With this procedure we obtained a lower density parameter
$\rho_c$ = 0.140 fm$^{-3}$ than the saturation density
$\rho_c$ = 0.160 fm$^{-3}$.

In practice, calculated charge-radii also
depend on the active pairing space.
In the present paper, 
we take into account all bound levels with energies
lower than $e_F$ + 1$\hbar \omega$, where $e_F$ is
the Fermi energy and 
1$\hbar\omega$ = 41A$^{-1/3}$ MeV.
With this prescription, however, we have encountered a problem of convergence.
In each step of iterations,
some of the highest levels go in and out from the active
pairing space, which give rise to an oscillation of the binding
energy. Thus for the first few hundred iterations,
we excluded the gap potential \D{q} from the mean field
hamiltonian and solved the equation by diagonalization.
After having had a pretty good convergence,
we fixed the number of single-particle levels of the active
pairing space and solved the full mean-field equation
by employing the gradient step.

We have also studied the cut-off prescription
of Bonche et al. {\cite{BF85}}.
With their prescription, however,
we could not reproduce odd-even staggering
in light Ba isotopes if we included
the positive evergy levels in the active pairing space.
Except for this difficulty,
our cut-off prescription gives similar results to those of
Bonche et al. {\cite{BF85}}.

We assumed 
$ {\cal D} = 1 / (E_{kin} + 50) $ MeV$^{-1}$
as a size of the gradient step.
The parameter ${\cal D}$ varies during the iterations
and $E_{kin}$ is the kinetic energy of the previous iteration.
We have also assumed that the iteration has converged 
if $|\Delta E| < $ 10$^{-4}$ MeV for the energy
and $|\Delta r| <$ 10$^{-4}$ fm for the charge radius
between two successive iterations.
(We did not have a good solution for $^{125}$Sn, 
where the charge radius fulfilled 
$|\Delta r| < $ 10$^{-4}$ fm,
but the binding energy oscillated with amplitude of
$|\Delta E| \simeq$  2 $\times$ 10$^{-3}$ MeV.)

Single particle states were expanded
in an deformed oscillator basis with axially-symmetry {\cite{V73}}. 
Eleven major shells were assumed in the calculation. 
Eleven major shells are
not enough to produce correct binding energies of 
heavy nuclei. 
However, by comparing the present results with
those of fifteen major shells, 
we have found that neutron separation energies and 
isotopic shifts were little affected by the size of the basis space. 
Odd neutron nuclei were treated in 
the blocking approximation {\cite{RS80}}. 
Odd neutron contributes to the density $\rho$, 
but it does not contribute to the pairing density \k{}.

Fig.~1 shows calculated results of Sn isotopes along
with experiment;
(a) neutron separation energies,  
(b) isotopic shifts in the charge radii
of Sn isotopes normalized to the nucleus $^{120}$Sn
($\Delta r^2_{c} = r^2_{c}(A) - r^2_{c}(120)$), and 
(c) isotopic shifts in the charge radii of Sn isotopes.
Experiment is denoted by open circles,
while calculation is denoted by crosses.
As a guide for eyes, they are connected with 
solid lines and dashed lines, respectively.
We also modified changes in the charge radii
of Fig.~1(b) by subtracting an equivalent of the liquid-drop
difference, $r_{LD}^2(A) = 3/5 r^2_0 A^{2/3}$ and $r_0$ = 1.2 fm.

Figs.~1(a) and 1(b) show good agreement with experiment
both for neutron separation energies and changes
in the charge radii of Sn isotopes.
Close examination of Fig.~1(b)
reveals that charge radii of odd neutron nuclei are
smaller than the average radii of their even neighbors.
This is a well-known phenomena of 
odd-even staggering of isotope shifts.
In Fig.~1(c) the phenomena is more clearly exhibited by
taking the difference of charge radii between 
neighboring isotopes.

Odd-even staggering of isotope shifts has been 
a longstanding problem in nuclear physics. 
Only recently it is realized {\cite{RZ88,FT94,KT95,FZ96,TB93}} 
that a density dependence seems to be necessary
for the pairing and that mean field effects of the pairing
seems to be responsible for this phonomena. 
Indeed, we cannot predict odd-even staggering 
of charge radii within a framework of ordinary
HF + BCS theory, even if we employ the density dependent pairing 
in the calculation.
We need to take into account
mean fields of the pairing 
to explain the phenomena.

Fig.~2(a) shows the HF + BCS calculation without
mean fields of the pairing.
As is mentioned above,
we could not reproduce odd-even staggering 
of isotope shifts even if
we employed the density dependent pairing
in the calculation. 
Fig.~2(b) shows the HF + BCS calculation with 
the pair mean-field \G but without the neutron gap-potential \D{n} .
Odd-even staggering is now well reproduced
except for heavy Sn isotopes.
Fig.~2(c) shows the HFB calculation with the gap potential
but without the pair mean-field.
The figure shows that the gap potential is also
responsible for the odd-even staggering.
The potential works especially well for heavy Sn isotopes. 
Reminding that Fig.~1(c) is the full HFB calculation
where the pair mean-field and the gap potential
are both included in the hamiltonian,
we may conclude that the pair mean-field as well as
the gap potential are both responsible for the
occurence of odd-even staggering of Sn isotopes.

Fig.~2(d) shows the pair mean-fields of 
$^{109}$Sn, $^{110}$Sn, and $^{111}$Sn.
Fig.~2(e) shows the neutron gap potentials of the same nuclei.
In these figures we show monopole radial shapes of the
potentials in the Legendre expantion.
From Fig.~2(d) we see that the pair mean-field is 
repulsive inside a nucleus.
The repulsive potential causes a shallow HF field
and thus the nucleus expands.
Since an even nucleus has a stronger 
pair mean-field than its odd neighbors,
the charge radius of the even nucleus expands
more than its odd neighbors.
The same is true for the gap potential.
Fig.~2(e) shows that the gap potential
is repulsive inside a nucleus while it is 
attractive at the nuclear surface.
The potential too, expands the nucleus.
The potential is also stronger for an even nucleus
than its odd neighbors, leading to a larger charge radius
for the even nucleus than its odd neighbors.

It is noted that the pair mean-field has
a larger effect than the gap potential on the odd-even staggering
of isotope shifts.
This is because that the pair mean-field \G 
exists in the proton HF field
even for proton closed-shell nuclei.
On the other hand, the proton gap-potential \D{p}
vanishes for the proton closed-shell nuclei.
For such nuclei, the neutron gap-potential \D{n} affects only
indirectly the charge radii of isotopes via the HF fields.

Fig.~3(a) shows one-neutron separation energies of Pb isotopes. 
Fig.~3(b) shows isotopic shifts in the charge radii 
of Pb isotopes normalized to the nucleus $^{208}$Pb.
Fig.~3(c) shows isotopic shifts in the charge radii of Pb isotopes.
Good agreement with experiment was obtained both for 
neutron speration energies and isotopic shifts.

Pb isotopes are similar to Sn isotopes 
in that the proton shell is closed. 
However, in contrast to Sn isotopes, 
the isotopic chain of Pb extends 
beyond the neutron shell closure at N=126. 
At the shell closure
we see in Fig.~3(b) an abrupt change of charge-radius differences,
which is called a ``kink''.

We have also investigated mean field effects of the pairing
on the kink of Pb isotopes.
Mean fields of the pairing get stronger as an isotope goes away
from $^{208}$Pb and strong mean fields could give rise to 
a larger charge radius.
If this is the case, changes in the charge radii of Pb isotopes
would be small for nuclei lighter than $^{208}$Pb,
while they would be large for nuclei heavier than $^{208}$Pb.
Fig.~3(b) shows isotopic shifts of Pb nuclei
from our calculation, the relativistic mean field (RMF)
theory {\cite{SL93}}, and experiment.
Our calculation could not predict large enough isotope shifts
for nuclei beyond $^{208}$Pb.
Indeed our calculation is very similar to 
the HF + BCS calculation without the mean field
effects of the pairing. Mean fields of the pairing seem to 
have only marginal effects on the occurence of the kink around $^{208}$Pb.

The kink around $^{208}$Pb may be primarily due to the
shell structures in this region of nuclei.
The 2g$_{9/2}$ neutron level lying close to the continuum
seems to be the key to the occurence of the kink.
Indeed the kink is well reproduced with 
the RMF calculation{\cite{SL93}}.
The kink is also reproduced with a nonrelativistic Skyrme calculation by 
omitting the Fock term of the spin-orbit interaction{\cite{SL95}}
or by assuming the isospin dependence
in the spin-orbit energy functional{\cite{SL95,RF95}}.

We have so far studied mean field effects of the pairing
on the charge radii of spherical nuclei.
In the following we will examine mean field effects of
the pairing on the charge radii of 
transitional and well deformed nuclei.

Fig.~4(a) shows neutron separation energies of Ba isotopes.
Fig.~4(b) shows isotopic shifts in the charge radii
of Ba isotopes. 
As in the case of Pb isotopes, the isotopic chain
extends beyond the shell closure
at N = 82 and exhibits a kink around $^{138}$Ba.
Fig.~4(c) shows odd-even staggering of Ba isotopes.
Agreement with experiment is in general good.

The kink in Ba isotopes may be due to 
changes in the nuclear deformation. 
Proton shells of Ba nuclei are not closed and 
therefore most Ba nuclei are soft against nuclear deformation.
Fig.~4(d) shows calculated quadrupole moments
which agree well with experiment {\cite{RM87}}.
As seen from this figure, the deformation 
decreases as the neutron number increases 
toward $^{138}$Ba and then the deformation increases beyond $^{138}$Ba.
Thus changes in the charge radii are
small for nuclei lighter than $^{138}$Ba,
while they are large for nuclei heavier than $^{138}$Ba.
We see in Fig.~4(b) that
our calculation reproduces the kink fairly well,
though the deviation from experiment becomes
appreciable in the region of light Ba isotopes.

At this point we wish to mention the reason 
why we have employed a small density parameter
\r{c} = 0.140 fm$^{-3}$ for the pairing 
instead of the saturation density \r{c} = 0.160 fm$^{-3}$.
Fig.~5(a) shows neutron separation energies of Ba isotopes
calculated with \r{c} = 0.160 fm$^{-3}$ and
$V_0$ = --1000 MeV$\cdot$fm$^{3}$.
(A strength of $V_0$ = -- 1000 MeV$\cdot$fm$^{3}$
is necessary in order to fit to neutron separation energies.) 
Fig.~5(b) shows isotopic shifts
between neighboring isotopes calculated with 
the same set of parameters.
By comparing Fig.~4(c) and Fig.~5(b),
we see that with this set of parameters
odd-even staggering of isotope shifts does not
agree well with experiment.
Although we have shown Ba isotopes as an example,
we have obtained good results
also for Sn and Pb isotopes with
\r{c} = 0.140 fm$^{-3}$ and $V_0$ = --1250 MeV$\cdot$fm$^{3}$.

In Fig.~6, we show neutron separation energies of Yb isotopes.
Agreement with experiment is good.
Figs.~6(b) and 6(c) show changes in the charge radii of Yb isotopes.
In this case, however, agreement with experiment is rather poor
especially for nuclei lighter than $^{165}$Yb.
Our calculation suggests that 
odd-even staggering in the charge radii of Yb isotopes
might be related to the odd-even staggering of nuclear 
deformation of these nuclei.
For nuclei lighter than $^{165}$Yb,
a larger deformation is calculated for an even isotope
than its odd neighbors.
The difference is enough to explain the odd-even sttagering
of charge radii of these nuclei.
For nuclei heavier than $^{165}$Yb,
however, a larger deformation is calculated for an odd nucleus
than its even neighbors.
Mean fields of the pairing and nuclear deformation 
have opposite effects on the charge radii,
leading to a rather complicated
behavier of isotopic shifts in this region of nuclei.

\section{Summary and Conclusion}{\label{summary}}

We have investigated mean field effects of 
the pairing on the charge radii of Sn, Pb, Ba, and Yb isotopes. 
There are two mean fields arising from the pairing energy functional.
One is the pair mean-field \G which comes from 
the density dependent part of the pairing. 
The pair mean-field does not depend on the charge state
and therefore the same form of the pair mean-field appears
in the HF fields of protons and neutrons.
In addition, the potential exists even for
proton closed-shell nuclei.
The other is the gap potential which arises from 
the variation of the pairing density in the pairing energy functional.
Therefore the gap potential arises also from 
the density independent pairing. 
In contrast to the pair mean field,
however, the proton gap potential vanishes for 
proton closed-shell nuclei.

For the pairing we have employed in the present calculations,
the pair mean-field is repulsive inside a nucleus. 
The repulsive potential causes a shallow HF field 
and the nucleus expands. 
Since an even nucleus has a stronger pair mean-field
than its odd neighbors, the charge radius of 
an even nucleus expands more than its odd neighbors. 
The same is true for the gap potential. 
It is repulsive inside a nucleus and 
attractive at the nuclear surface. 
The potential too, expands the nucleus. 
Like the pair mean field, the gap potential is
stronger for an even nucleus than its odd neighbors. 
These facts seem to be the key ingredient in the occurence of
odd-even staggering of isotopic shifts.

The present analyses have shown that the lower density parameter 
$\rho_c$=0.140 fm$^{-3}$ than the saturation density 
$\rho_c$=0.160 fm$^{-3}$ gives better agreement with experiment
in fitting to the odd-even staggering of isotope shifts of 
Sn, Pb, Ba, and Yb. 
We have also shown that the kink observed in the charge radii of Ba
isotopes may be due to changes in the nuclear deformation. 
The effects are enough to explain 
the kink around $^{138}$Ba.

\section*{ACKNOWLEDMENTS}

We would like to express our sincere appreciation to 
Ken-ichiro Arita for many fruitful and critical discussions. 
We would like to thank 
M. Yamagami and K. Matsuyanagi, and also members 
of Nagoya Nuclear-Structure Seminor for many illuminating
discussions.

\begin{figure}[htbp]
\epsfxsize=.45\textwidth
\centerline{\epsffile{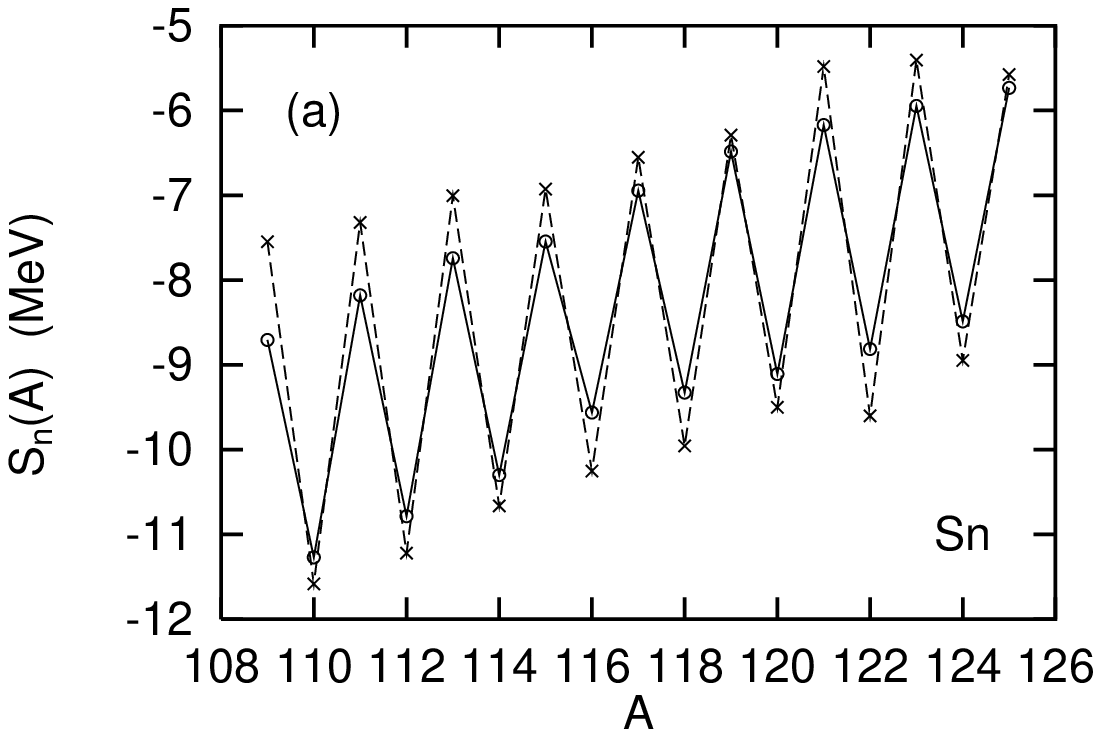}}

\begin{minipage}{.45\textwidth}
\epsfxsize=\textwidth
\epsffile{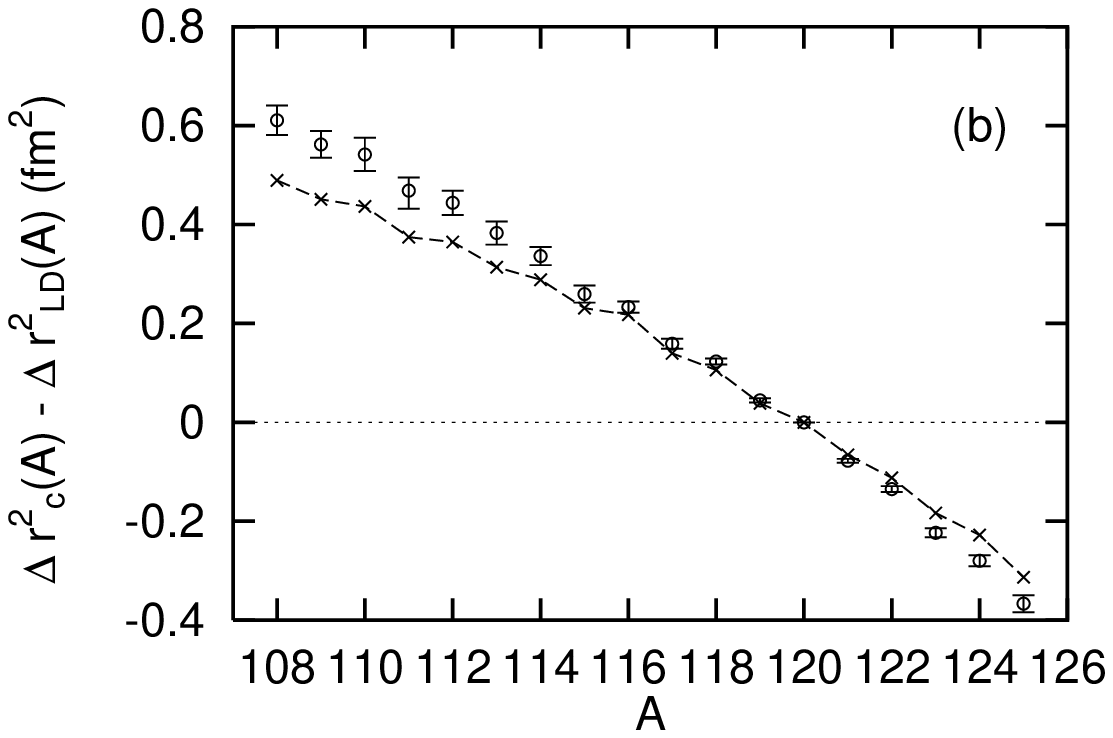}
\end{minipage}
\begin{minipage}{.45\textwidth}
\epsfxsize=\textwidth
\epsffile{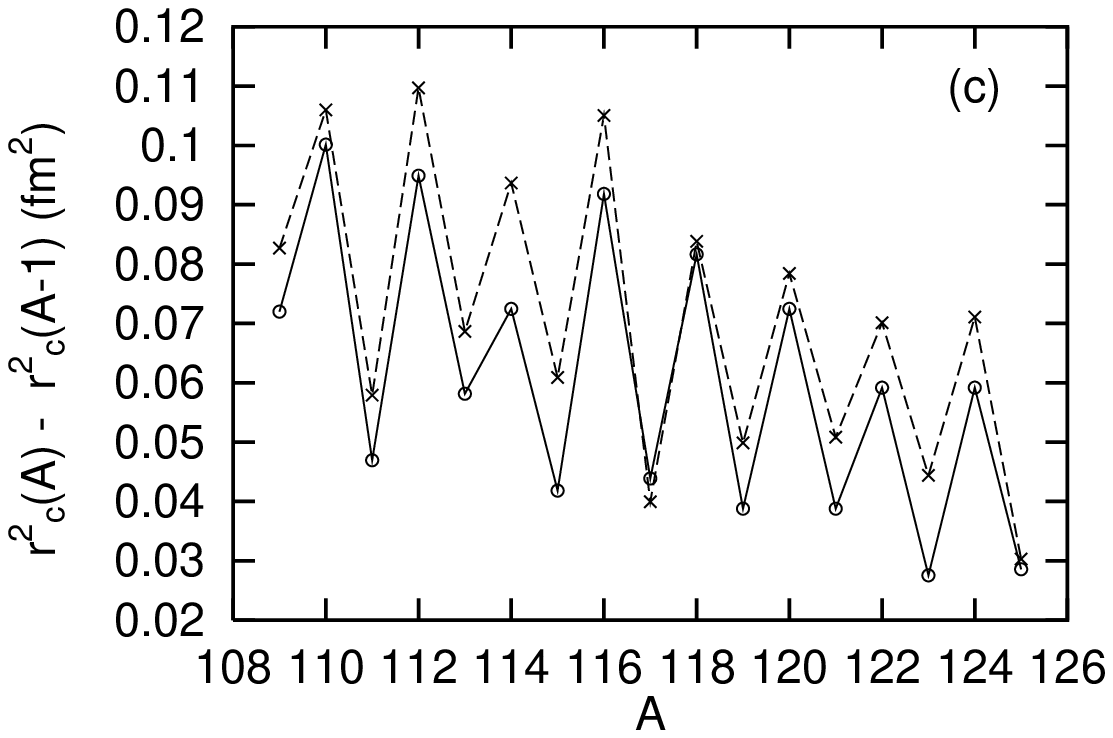}
\end{minipage}
\caption{\label{Sn1}
(a) Neutron separation energies of Sn isotopes. 
(b) Isotopic shifts in the charge radii of Sn isotopes 
normalizaed to the nucleus $^{120}$Sn. 
They are subtracted by an equivalent of the liquid-drop difference. 
(c) Isotopic shifts in the charge radii of Sn isotopes. 
Experiment is denoted by open circles, 
while calculation is denoted by crosses. 
As a guide for eyes, they are connected with 
solid lines and dashed lines, respectively. 
Experimental data are taken from
Refs. \protect \cite{AH87,AW93,ED87}.
}
\end{figure}
\newpage
\begin{figure}[htbp]
\epsfxsize=.45\textwidth
\centerline{\epsffile{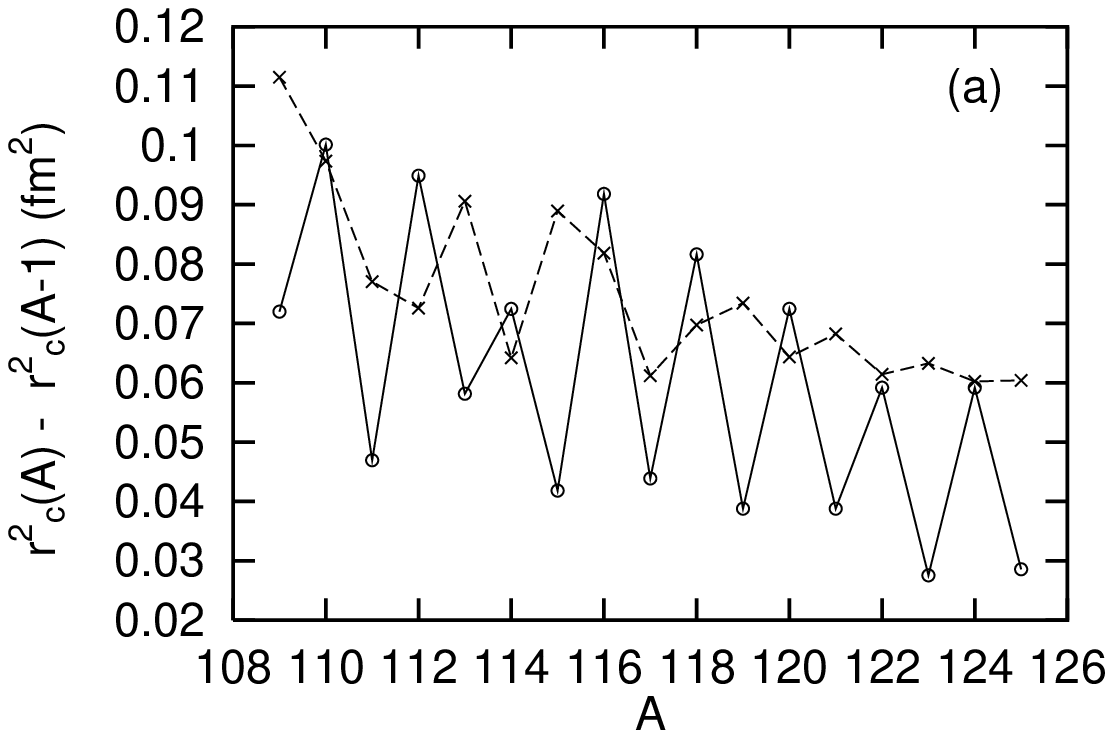}}

\begin{minipage}{.45\textwidth}
\epsfxsize=\textwidth
\epsffile{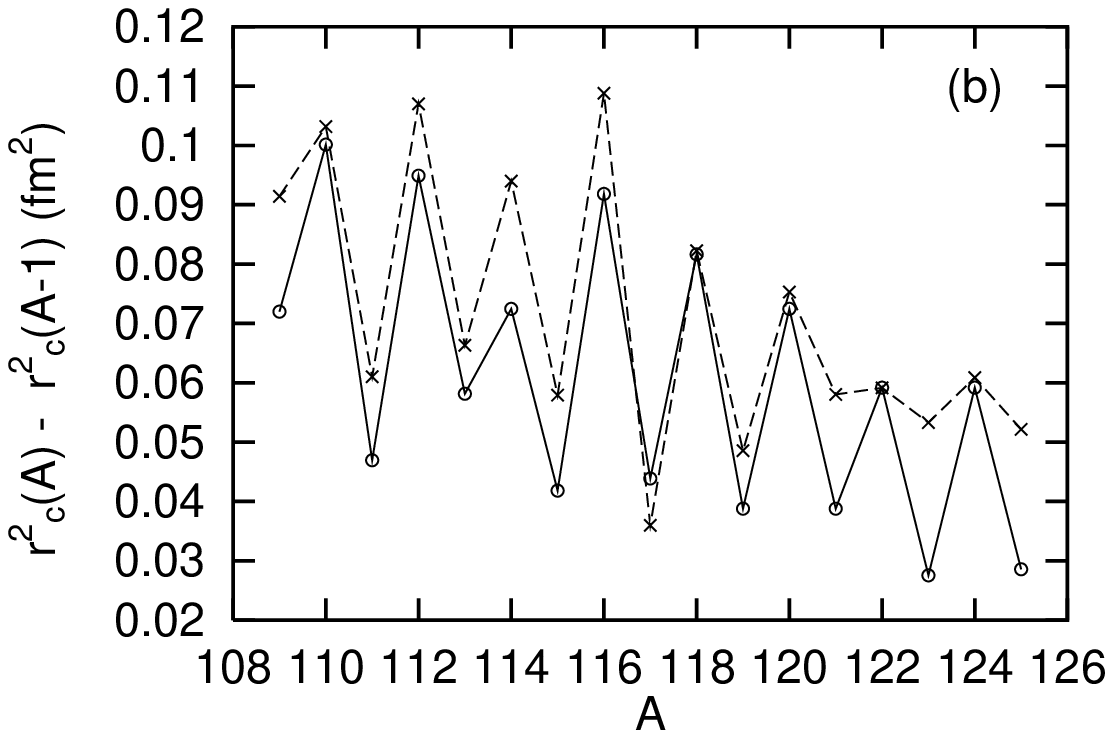}
\end{minipage}
\begin{minipage}{.45\textwidth}
\epsfxsize=\textwidth
\epsffile{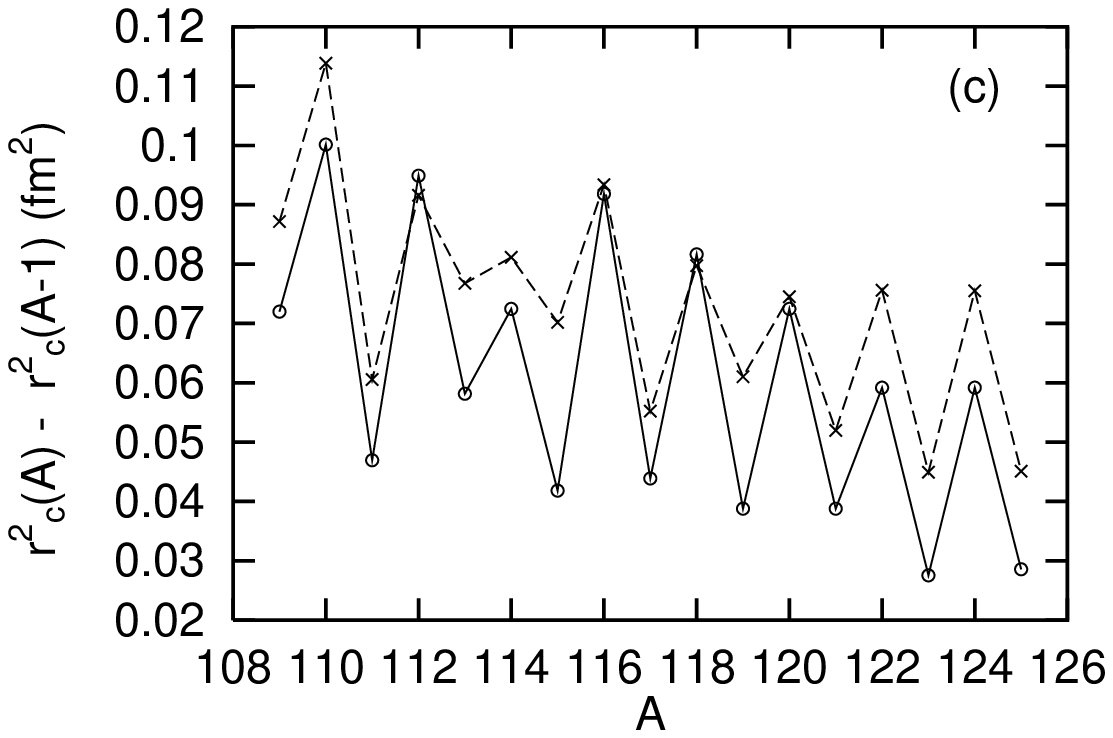}
\end{minipage}

\begin{minipage}{.45\textwidth}
\epsfxsize=\textwidth
\epsffile{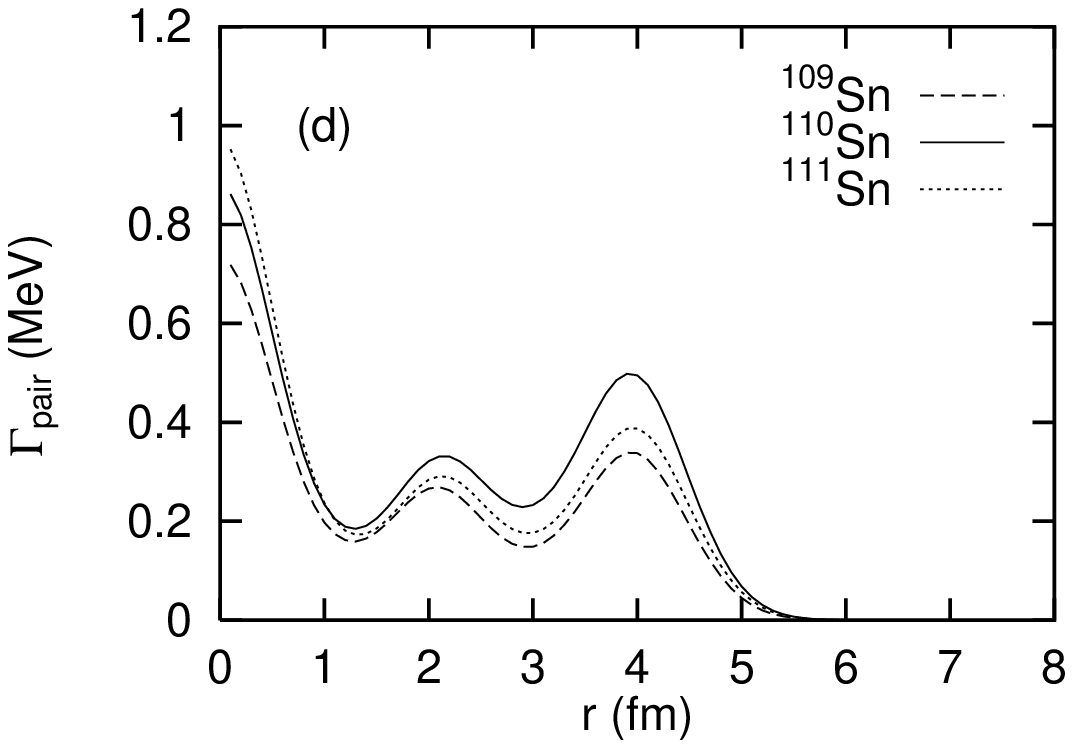}
\end{minipage}
\begin{minipage}{.45\textwidth}
\epsfxsize=\textwidth
\epsffile{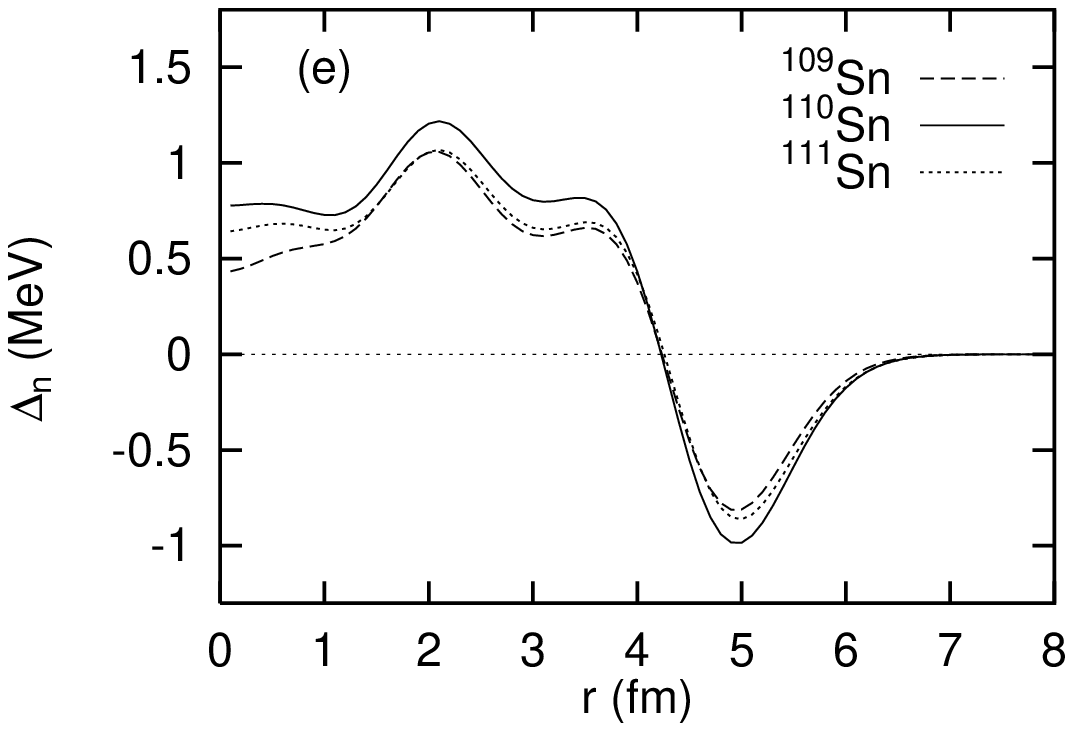}
\end{minipage}
\caption{\label{Sn2} 
Mean field effects of the pairing on the isotopic shifts of Sn. 
(a) The HF+BCS calculation without the mean fields of the pairing. 
(b) The HF+BCS calculation with the pair mean-field \G 
but without the neutron gap-potential \D{n}. 
(c) The HFB calculation with the neutron gap potential but without 
the pair mean-field. 
Experiment is denoted by open circles, 
while calculations are denoted by crosses.
(d) Monopole radial shapes of the pair mean-fields. 
(e) Monopole radial shapes of the neutron gap potentials.
}
\end{figure}

\begin{figure}[htbp]
\epsfxsize=.45\textwidth
\centerline{\epsffile{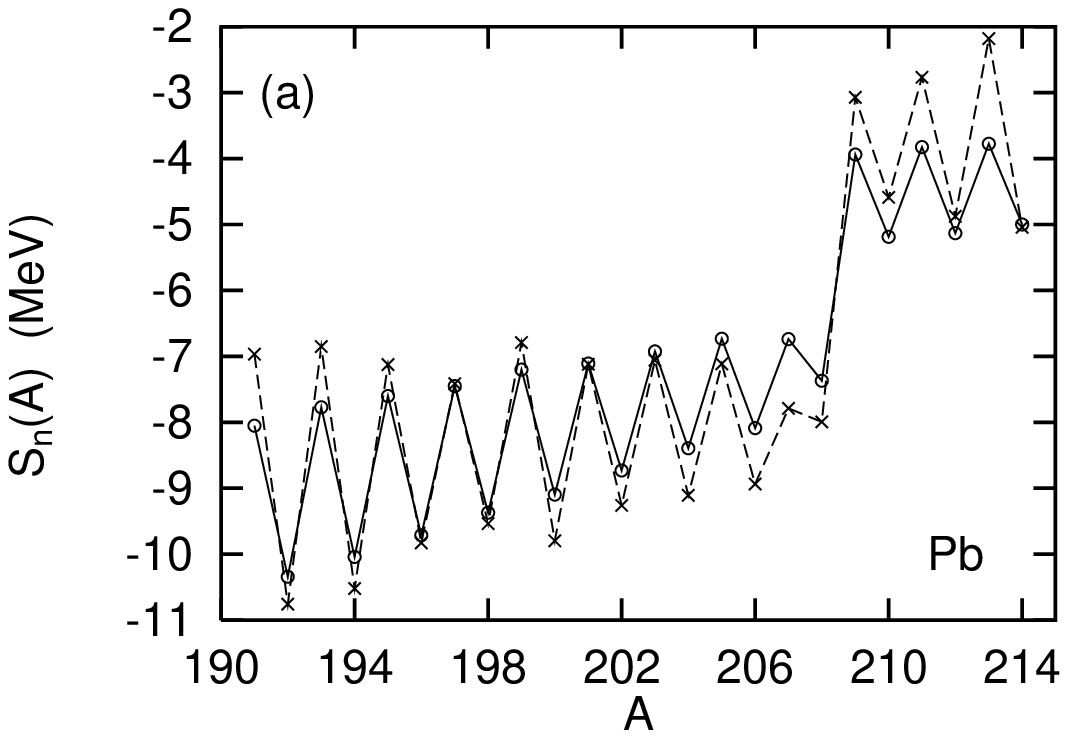}}
\begin{minipage}{.45\textwidth}
\epsfxsize=\textwidth
\epsffile{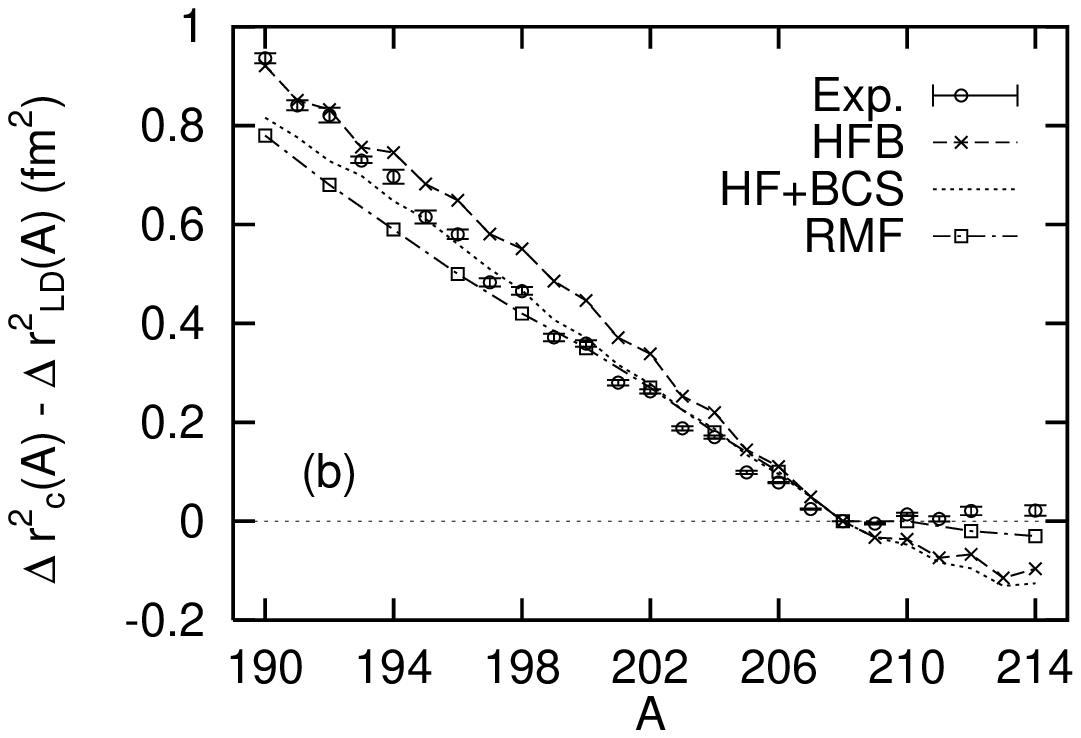}
\end{minipage}
\begin{minipage}{.45\textwidth}
\epsfxsize=\textwidth
\epsffile{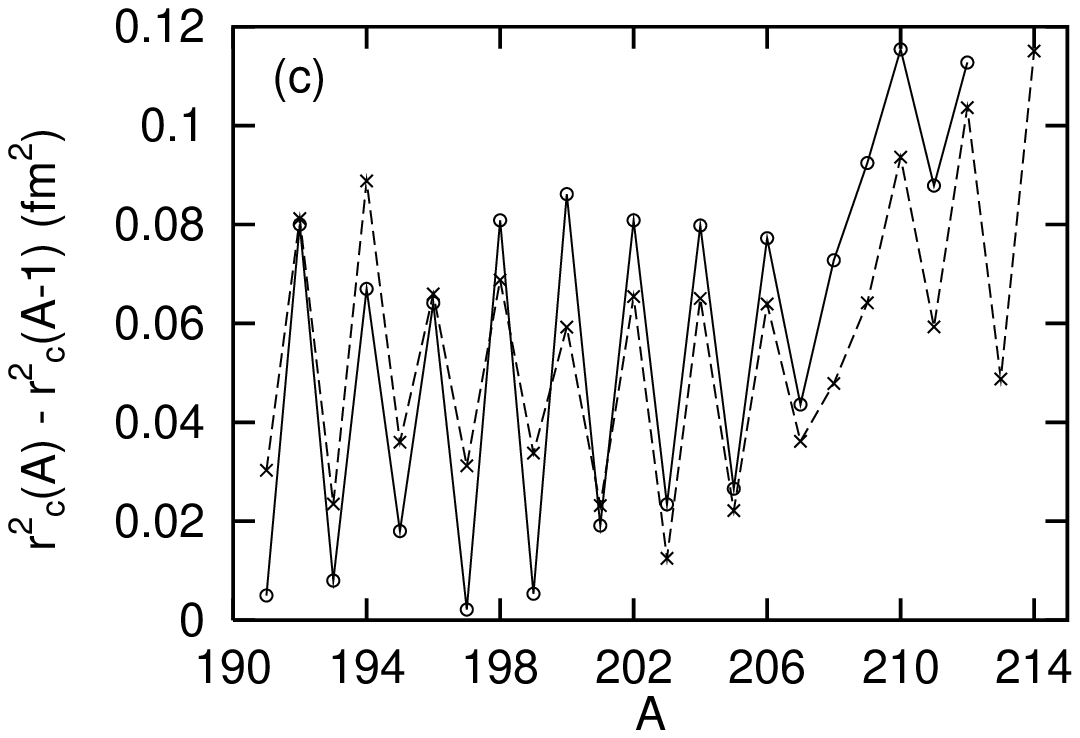}
\end{minipage}
\caption{\label{Pb1} 
(a) Neutron separation energies of Pb isotopes. 
(b) Isotopic shifts in the charge radii of Pb isotopes 
normalized to the nucleus $^{208}$Pb. 
The RMF calculation of Sharma et al. \protect \cite{SL93}
and the HF + BCS calculation are also shown for comparison. 
(c) Isotopic shifts in the charge radii of Pb isotopes. 
Experiment is denoted by open circles, 
while calculation is denoted by crosses.
Experimental data are taken from
Refs. \protect \cite{AH87,AW93,DE87,DK91}.
}
\end{figure}

\begin{figure}[htbp]
\begin{minipage}{.45\textwidth}
\epsfxsize=\textwidth
\epsffile{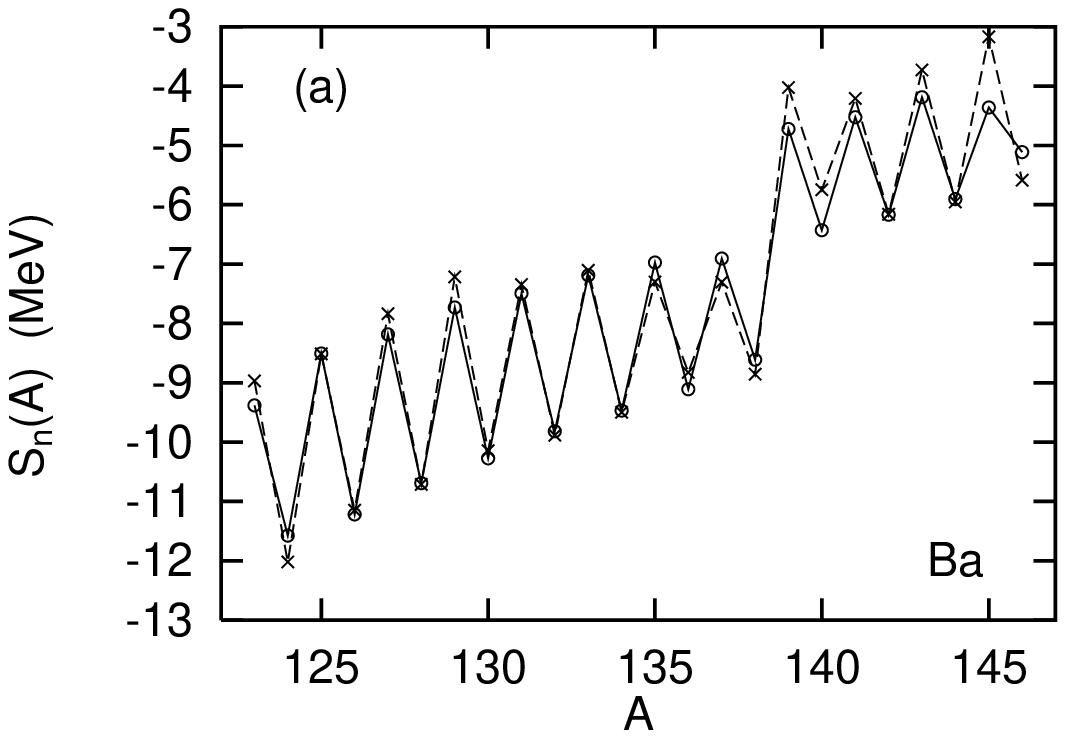}
\epsfxsize=\textwidth
\epsffile{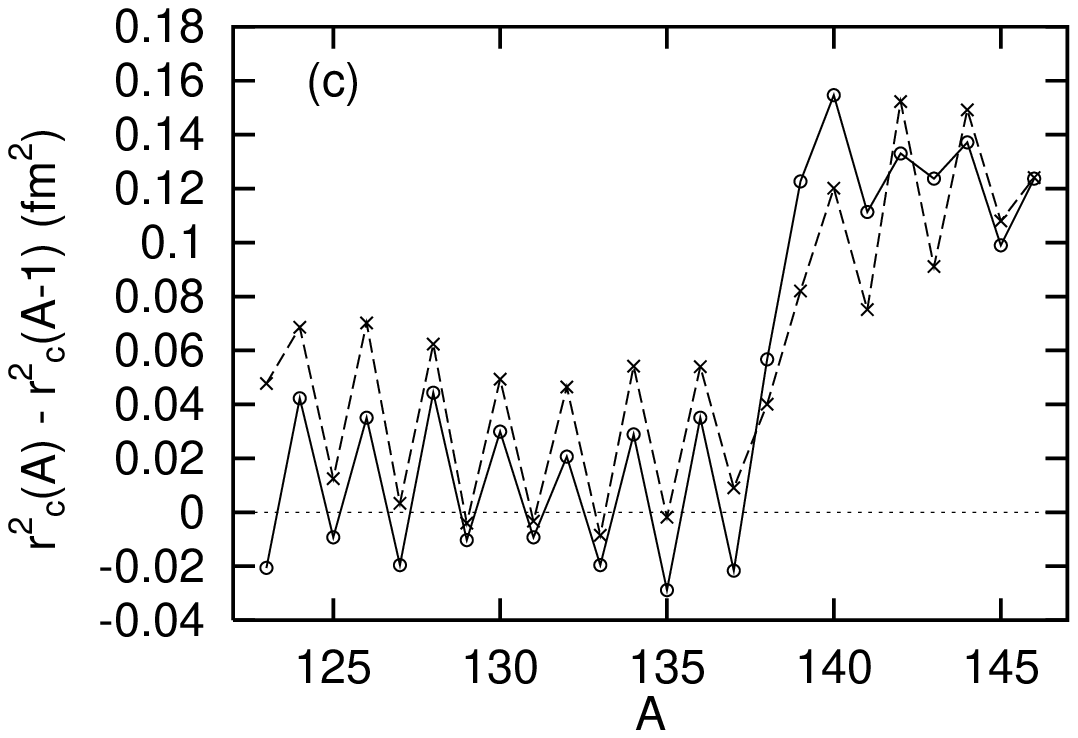}
\end{minipage}
\begin{minipage}{.45\textwidth}
\epsfxsize=\textwidth
\epsffile{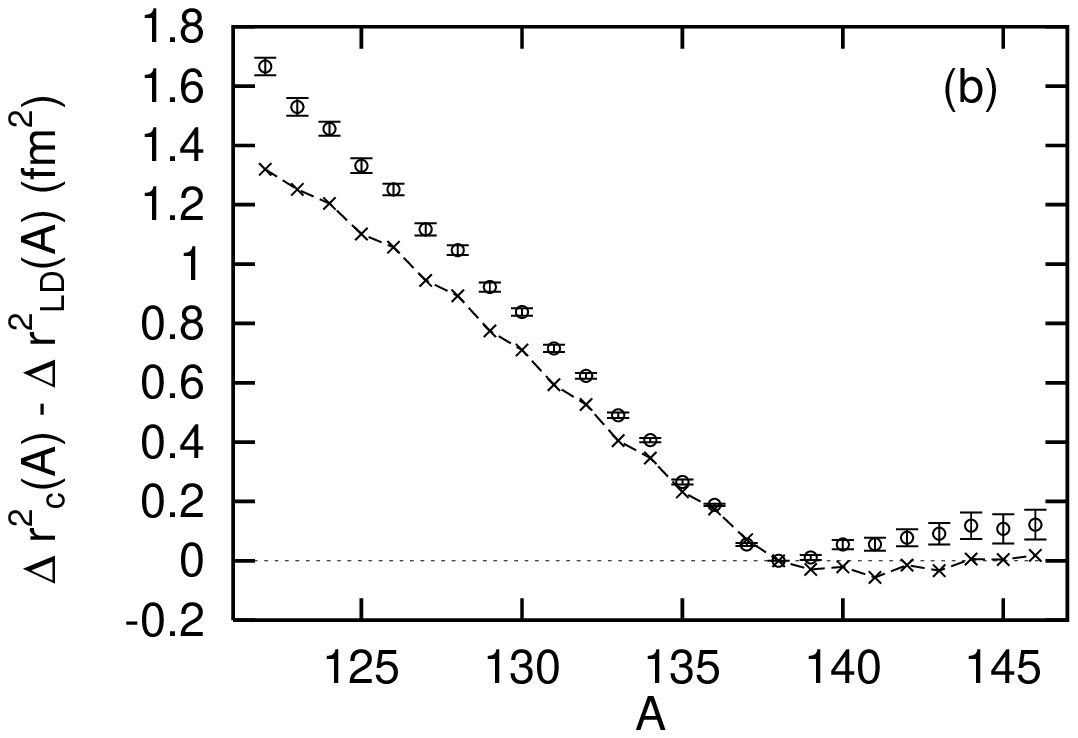}
\epsfxsize=\textwidth
\epsffile{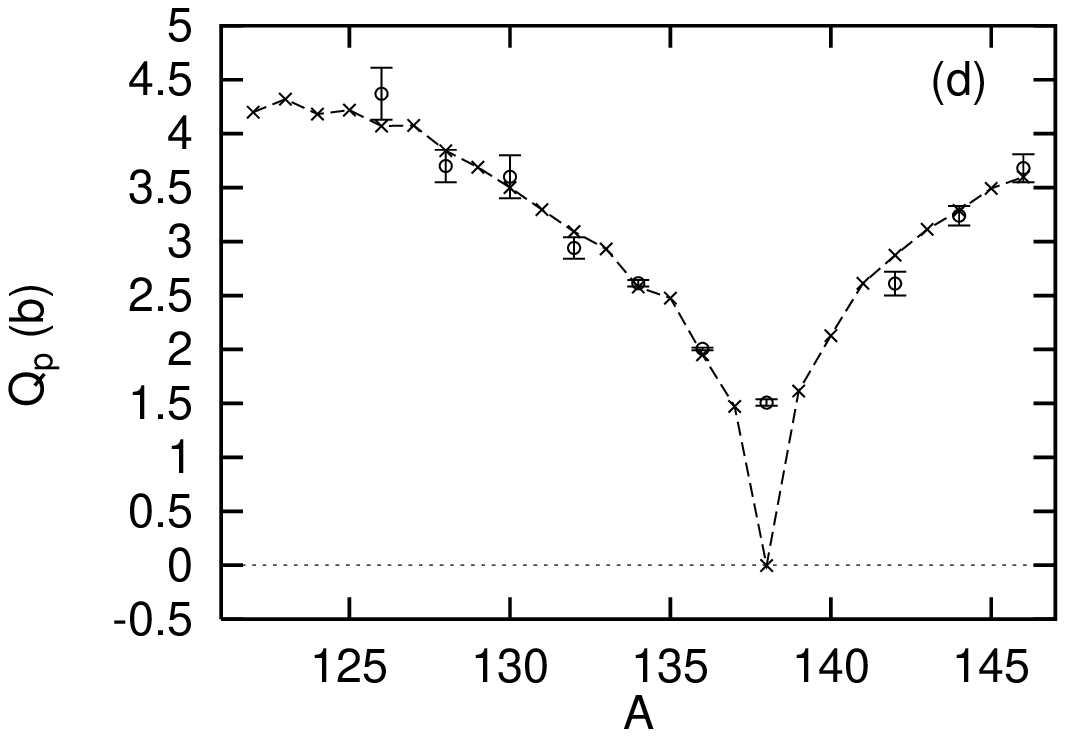}
\end{minipage}
\caption{\label{Ba1} 
(a) Neutron separation energies of Ba isotopes. 
(b) Isotopic shifts in the charge radii of Ba isotopes 
normalized to the nucleus $^{138}$Ba. 
(c) Isotopic shifts in the charge radii of Ba isotopes.  
(d) Intrinsic quadrupole moments of Ba isotopes. 
Experiment is denoted by open circles, 
while calculation is denoted by crosses.
Experimental data are taken from Refs.
\protect \cite{AH87,AW93,RM87}.
}
\end{figure}

\begin{figure}[htbp]
\begin{minipage}{.45\textwidth}
\epsfxsize=\textwidth
\epsffile{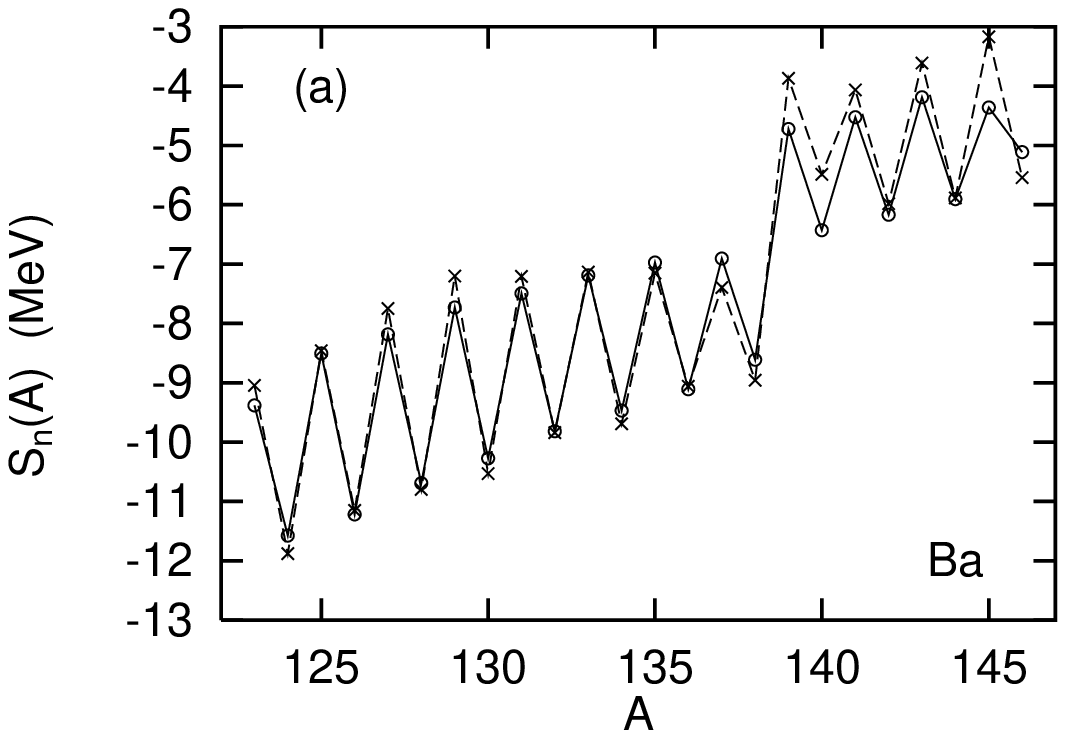}
\end{minipage}
\begin{minipage}{.45\textwidth}
\epsfxsize=\textwidth
\epsffile{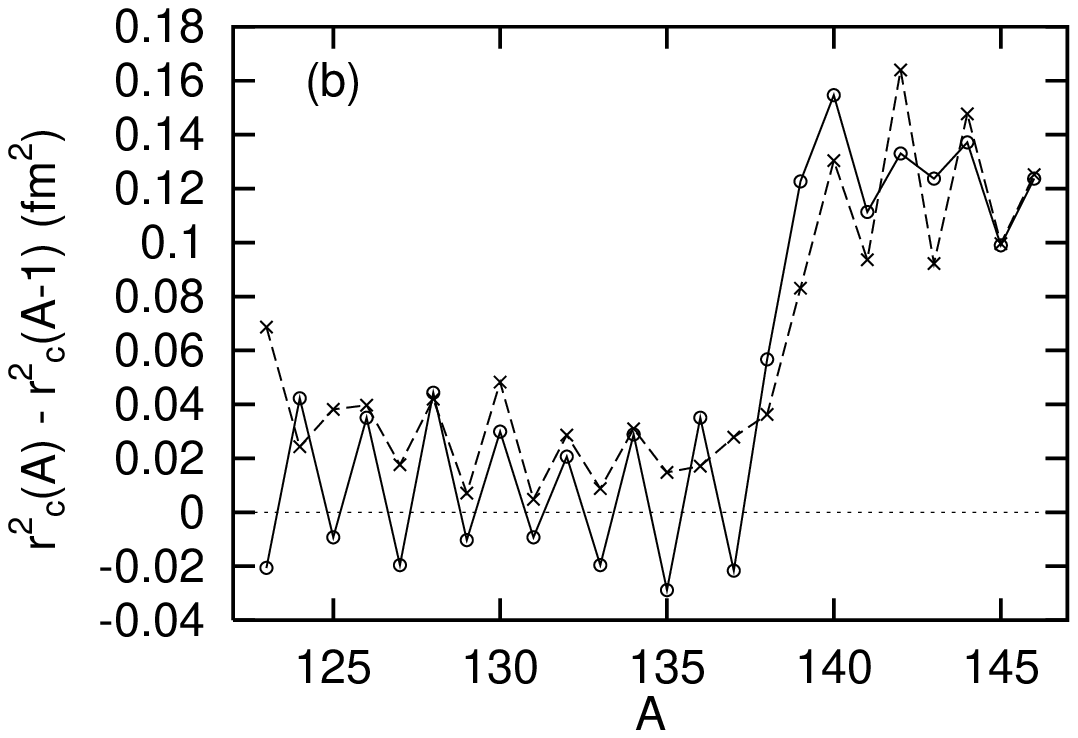}
\end{minipage}
\caption{\label{Ba2} 
(a) Neutron separation energies of Ba isotopes
calculated with the pairing parameters of 
$\rho_c$=0.160 fm$^{-3}$ and $V_0$= -- 1000 MeV$\cdot$fm$^{3}$.
(b) Isotopic shifts in the charge radii of Ba isotopes calculated 
with the same set of pairing parameters. 
Experimental data are taken from Refs.\protect \cite{AH87,AW93}.
} 
\end{figure}

\begin{figure}[htbp]
\epsfxsize=.45\textwidth
\centerline{\epsffile{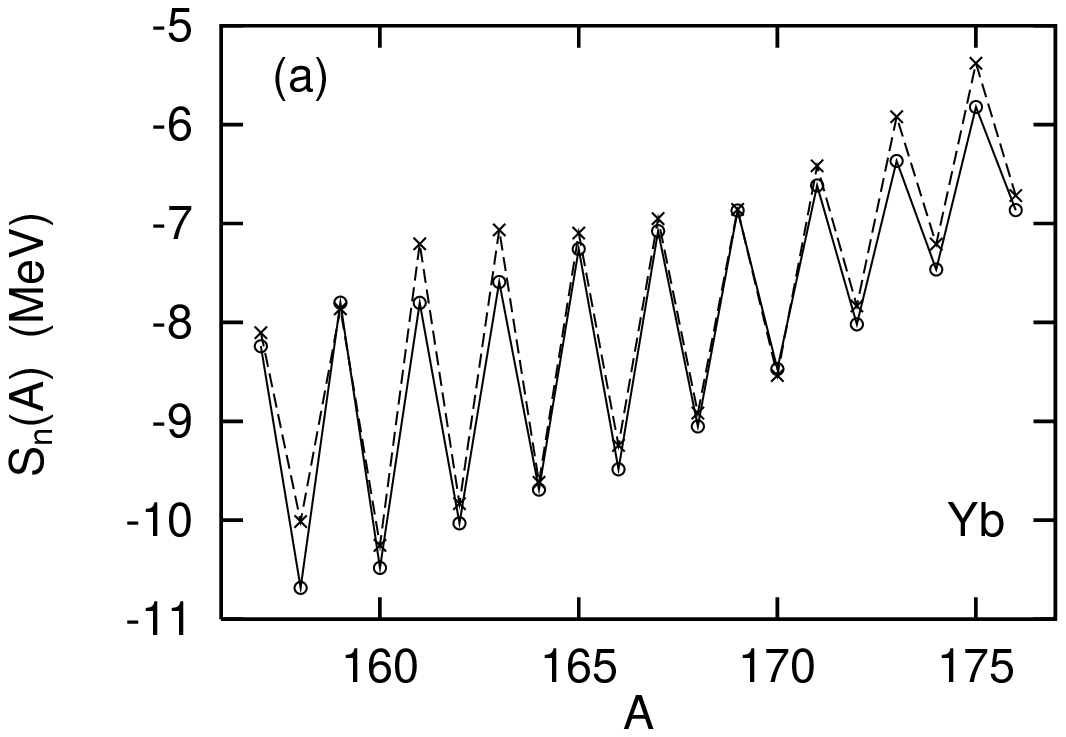}}

\begin{minipage}{.45\textwidth}
\epsfxsize=\textwidth
\epsffile{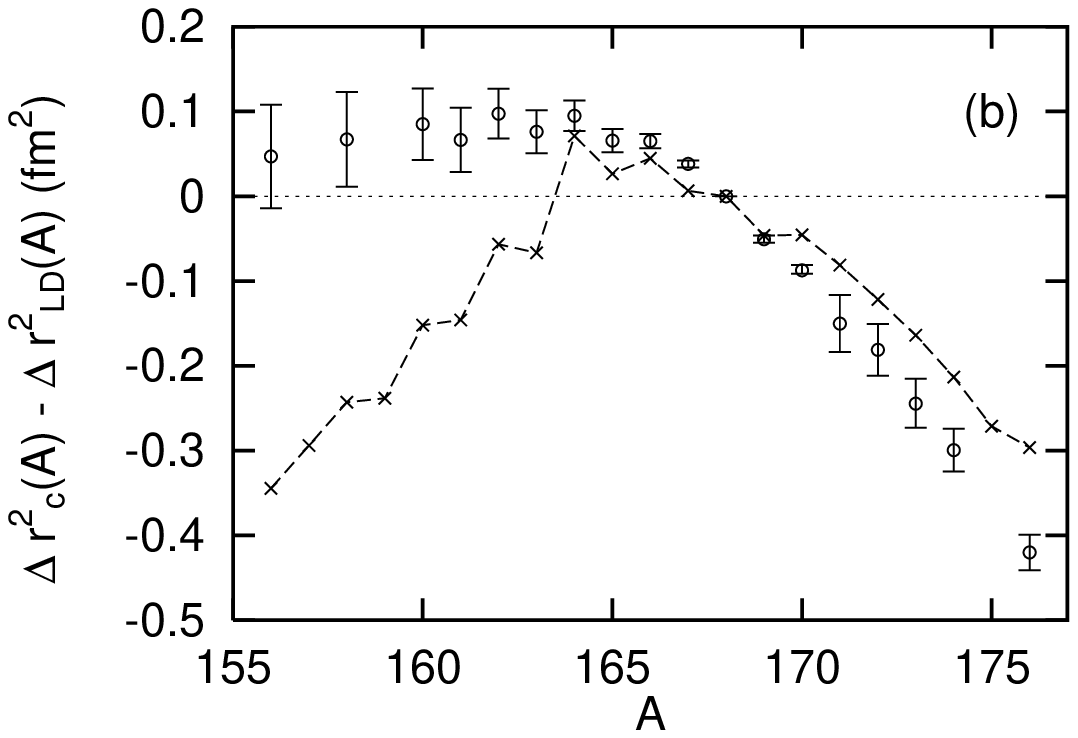}
\end{minipage}
\begin{minipage}{.45\textwidth}
\epsfxsize=\textwidth
\epsffile{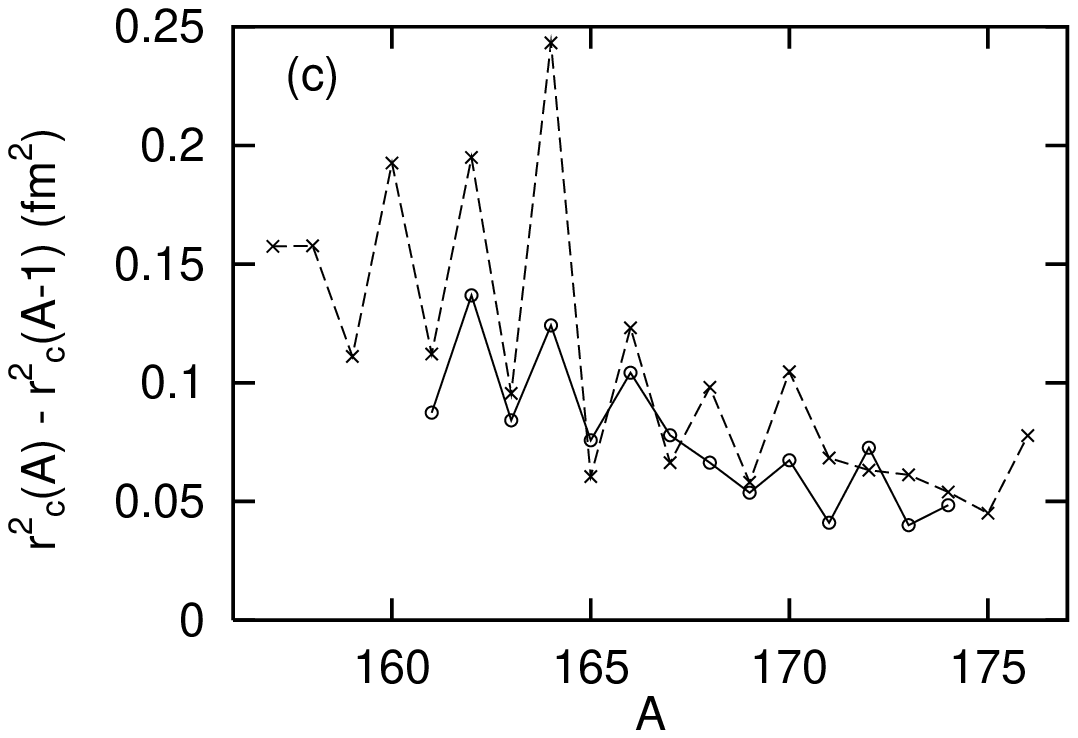}
\end{minipage}
\caption{\label{Yb1} 
(a) Neutron separation energies of Yb isotopes. 
(b) Isotopic shifts in the charge radii of Yb isotopes 
normalized to the nucleus $^{168}$Yb.  
(c) Isotopic shifts in the charge radii of Yb isotopes. 
Experiment is denoted by open circles, 
while calculation is denoted by crosses.
Experimental data are taken form Refs. 
\protect \cite{AH87,AW93} .
}
\end{figure}

\end{document}